\documentclass[intlimits,twoside,a4paper]{article}
\usepackage[cp1251]{inputenc}

\usepackage{cmpj3}

\issue{2018}{21}{2}{23703}
\doinumber{10.5488/CMP.21.23703}

\title[Anomalous zero-temperature magnetopolaronic blockade]{Anomalous zero-temperature magnetopolaronic blockade of resonant electron tunneling in Majorana-resonant-level single-electron transistor}

\author[G.A. Skorobagatko]{G.A. Skorobagatko\thanks{E-mail: gleb.a.skor@gmail.com}}

\address{Institute for Condensed Matter Physics of the National Academy of Sciences of Ukraine,\\
1  Svientsitskii St., 79011 Lviv, Ukraine }

\date{Received May 2, 2018, in final form June 18, 2018}

\begin{document}
\maketitle

\begin{abstract}
The magnetopolaronic generalization of a Majorana-resonant-level (MRL) model is considered for a single-level vibrating quantum dot coupled to two half-infinite $g=1/2$ Tomonaga-Luttinger liquid (TLL) leads. A qualitatively new non-trivial formula for the effective transmission coefficient and differential conductance for resonant magnetopolaron-assisted tunneling is obtained under the assumption about a fermion-boson factorization of corresponding averages. This approach is valid for the case of weak magnetopolaronic coupling in a system. Surprisingly, it is found that despite a supposed weakness of interaction between fermionic and bosonic subsystems in that case, a strongly correlated electron transport in the system reveals features of strong (and, hence, \textit{anomalous}) magnetopolaronic blockade at zero temperature if the energy of a vibrational quantum is the smallest (but nonzero) energy parameter in the system. Such an effect should be referred to as magnetic phase-coherent magnetopolaron-assisted resonant tunneling of Andreev type, that originates from a special, Majorana-like, symmetry of magnetopolaron-coupled tunnel Hamiltonian. The effect predicted in this paper can be used as an experimental fingerprint of Majorana-resonant level situation in single-electron transistors as well as for detection of ultra-slow zero-point oscillations of suspended carbon nanotubes in the Majorana-resonant level regime of electron tunneling through  corresponding single-electron transistors.

 
\pacs 73.63.Kv, 72.10.Pm, 73.23.-b, 73.63-b,71.38.-k, 85.85.+j
\end{abstract}

Resonant tunneling in strongly interacting electron systems, particularly, in different types of molecular single-electron transistors (SETs) is an attractive point in quantum mesoscopics due to the rich physics rooted in many related challenging problems  \cite{PP,PT,NR,GR,GS,KF,FL,MCK,KG1,KG2,NEW,MY}. The quantum mesoscopic setup in question is a single-electron molecular transistor modelled as a single-level quantum dot (QD) vibrating along the $y$-axis in a perpendicular (i.e., oriented along $z$-axis) constant magnetic field  \cite{OUR1,Pist}. Such a QD is weakly coupled to two one-dimensional (in $x$-direction) leads (long enough quantum wires or carbon nanotubes) by means of two tunnel barriers \cite{KG1}. The half-infinite one-dimensional leads imply an electron-electron interaction, which is described by Tomonaga-Luttinger liquid (TLL) model with conventional TLL correlation parameter $g=(1+U_{\text{TLL}}/\piup v_{\text F})^{-1/2}$, ($0<g<1$) defined by the ``bare'' constant $U_{\text{TLL}}$ of electron-electron interaction in TLL leads \cite{KR,AK}. For the most general situation of arbitrary $g$ and arbitrary magnetopolaronic coupling in a quantum dot, it is impossible to solve the electron transport problem exactly \cite{KR}. At the same time, in the most simple case of noninteracting (Fermi-liquid or FL) leads (if $g=1$) both polaronic and magnetopolaronic SET models behave very similarly and it is even difficult to distinguish between them \cite{OUR1, Pist}. On the other hand, as it was shown earlier for the special value $g=1/2$ of TLL correlation parameter, in the absence of any quantum vibrations of QD, the problem of resonant electron tunneling is exactly solvable even in the case of asymmetric tunnel coupling \cite{KF,KG1,NG}. In this model, in the case of symmetric tunnel couplings, the Majorana-like symmetry emerges in the tunneling Hamiltonian, and this is known as the TLL-realization of the Majorana-resonant-level model (MRLM) or simply as a spinless Tomonaga-Luttinger liquid resonant-level (TLLRL) model \cite{AK}. Recently, it was shown by the author for the polaronic generalization of TLLRL-model (see \cite{MY}) that in the case of strong electromechanical coupling, a novel type of Andreev-like resonant polaron-assisted tunneling is realized in the system. The latter polaronic $ g=1/2$ TLLRL model strongly differs from the polaronic model of SET with noninteracting (i.e., Fermi liquid) leads \cite{MY,NEW}. Thus, one may ask whether the magnetopolaronic MRL-model differs from polaronic MRL model or, as it takes place in the Fermi liquid case, these models are qualitatively similar? An unexpected answer to this question will be given herein below. 

In this paper, it is originally shown for the SET model with weak magnetopolaronic coupling that even in the case of symmetrical tunnel couplings between vibrating quantum dot and $g=1/2$ TLL leads, at the Toulouse point in Coulomb interaction strength (between TLL-leads and quantum dot) i.e., in the MRLM case \cite{AK,KG1}, the model exhibits a strong suppression of differential conductance in the zero-temperature limit of  electron transport through the junction. A corresponding novel formula for the transmission coefficient of strongly correlated electrons in the magnetopolaronic Majorana-resonant-level model is derived. 

It is reasonable to start with the Hamiltonian of magnetopolaronic $g=1/2$-TLLRL model  already in its re-fermionized form (one can see corresponding unitary transformations in references~\cite{MY,KG1}):
\begin{equation} \label{1}
\hat{H}=\hat{H}_{\text l}+\hat{H}_{\text d}+\hat{H}_{\text t}.
\end{equation}
Here, the first term describes a quadratic Hamiltonian of half-infinite one-dimensional $g=1/2$ TLL leads: $\hat{H}_{\text l}=\sum_{\pm}1/2\piup\int \rd x[\hat{\Psi}^{+}_{\pm}\hat{\Psi}_{\pm}(x)]^{2}=\sum_{\pm}1/2\piup\int \rd x[\partial_{x}\Phi_{\pm}(x)]^{2}$. (Here and below, we put $e=1$ and $\hbar v_{g}=\hbar v_{\text F}/g=1$ with ``bare'' Fermi velocity $v_{\text F}$.) Operators $\hat{\Psi}_{\pm}(x)=\exp[\ri\Phi_{\pm}(x)\sqrt{2}]/\sqrt{2\piup a_{0}}$ (here $g=1/2$) stand for \textit{composite} fermions (being spatially nonlocal in $x$-direction) (see e.g., reference~\cite{KG1} for details) and, hence, they fulfil standard fermionic anticommutation relations $\{\hat{\Psi}_{\pm}(x),\hat{\Psi}_{\pm}^{+}(x')\}=\delta(x-x')$ \cite{KG1}. The corresponding transformed bosonic phase fields $\Phi_{\pm}(x)$ are connected with conventional bosonic phase fields $\Phi_{j}(x)$ ($j=\text{L},\text{R}$) of plasmonic charge density excitations in the $j$-th lead ($j=\text{L,\,R}$) by relation: $\Phi_{\pm}(x)=[\Phi_{\text L}(x)\pm\Phi_{\text R}(x)]/\sqrt{2}$ (see reference~\cite{KG1}), where each $j$-th phase field is defined on the full axis $x \in (-\infty; +\infty)$ with its half-axis $(-\infty;-0)$ and $(+0;+\infty)$ corresponding to right- and left-moving chiral components of this quantum field. 

The difference between chemical potentials of noninteracting electrons in the remote reservoirs is proportional to the bias voltage $V$ being applied to the leads at the points $x=\pm \infty$ (see, \cite{KG1}). The second term in equation~(\ref{1}) $\hat{H}_{\text d}=\Delta\hat{d}^{+}\hat{d}+\frac{\hbar\omega_{0}}{2}(\hat{p}_{y}^{2}+\hat{y}^{2})$ represents the transformed Hamiltonian of a single-level vibrating quantum dot (QD) at the Toulouse point in Coulomb interaction between QD and TLL leads (see references~\cite{KG1,KG2}). Here, $\hat{d}^{+}\,(\hat{d})$ are  standard fermionic creation (annihilation) operators at a resonant level of QD; $\Delta=\Delta(V_{g})$ is the resonant level energy driven by $V_{g}$-gate voltage being applied to QD electrostatically.\footnote{All energies in the system are counted from the Fermi energy of TLL leads; in our model it is convenient to put Fermi energy to be equal to zero. Besides that, the ``magnetopolaronic'' shift of resonant level due to magnetopolaronic coupling is already taken into account in the parameter $\Delta(V_{g})$ \cite{MY,OUR1}.}  In the case of MRL-coupling, one should put $\Delta(V_{g})=0$. One can always satisfy the latter ``resonance'' condition since $ V_{g} $ is an independent parameter in the model. Thus, in the magnetopolaronic MRLM case of interest, one can write down  
\begin{equation} \label{2}
\hat{H}_{\text d}=\hat{H}_{\text v}=\frac{\hbar\omega_{0}}{2}\big(\hat{p}_{y}^{2}+\hat{y}^{2}\big),
\end{equation}
i.e., the transformed Hamiltonian of fermionic resonant level of QD in the magnetopolaronic MRLM case contains only vibronic degrees of freedom. In equation~(\ref{2}), $\hbar\omega_{0}$ is the energy of vibrational quantum, and $\hat{p}_{y}$, $\hat{y}$ are the dimensionless bosonic operators of the momentum and center-of-mass coordinate of QD in $y$-direction. 
Then, the transformed tunnel Hamiltonian [the third term in equation~(\ref{1})] takes the form:
\begin{eqnarray}\label{3}
\hat{H}_{\text t}=\hat{d}^{+}\left[\gamma_{\text L}\hat{X}_{\text L}^{+} \hat{\Psi}_{-}(0)+\gamma_{\text R}\hat{X}_{\text R}^{+} \hat{\Psi}_{-}^{+}(0)\right]
+\left[\gamma_{\text L} \hat{\Psi}_{-}^{+}(0)\hat{X}_{\text L}+\gamma_{\text R} \hat{\Psi}_{-}(0)\hat{X}_{\text R}\right]\hat{d},
\end{eqnarray}
where, following the references~\cite{KG1,KG2} we defined $\hat{\Psi}_{-}(0)=[\hat{\Psi}_{-}(0^{-})+\hat{\Psi}_{-}(0^{+})]/2$, and $\gamma_{\text{L(R)}}$ are the tunneling amplitudes for the left- and right TLL lead, correspondingly. (In our notations: $\gamma_{\text L}^{2}+\gamma_{\text R}^{2}=\Gamma_{0}$ is the ``bare'' width of the fermionic level of QD in a standard wide-band-limit (WBL) approximation \cite{KG1}.) 
A novel element in equation~(\ref{3}), as compared to the conventional exactly solvable $ g=1/2 $ TLLRL model of reference~\cite{KG1} without quantum vibrations, is the ``magnetopolaronic'' renormalization \cite{OUR1} of the tunneling amplitudes by bosonic operators:
\begin{equation} \label{4}
\left\{    \begin{array}{ll}
\hat{X}_{\text L}=\exp(-\ri\phi\hat{y}),  \\
\hat{X}_{\text R}=\exp(\ri\phi\hat{y}),
\end{array}
\right.
\end{equation}
that describe the influence of a fluctuating Aharonov-Bohm phase acquired by the electron in the process of resonant tunneling \cite{OUR1}. In equation~(\ref{4}), $\phi=\Phi/\sqrt{2}\Phi_{0}=e y_{0}D_{0}B_{z}/\sqrt{2}hc$ is a dimensionless magnetopolaronic coupling constant, where $y_{0}=\sqrt{\hbar/M\omega_{0}}$ is an amplitude of zero-point oscillations of the QD center-of-mass coordinate in the  $y$ direction ($M$ is the mass of quantum dot); $D_{0}$ is the characteristic distance between two TLL leads (i.e., the characteristic size of QD region in the $x$ direction); $B_{z}$ is the absolute value of $ z $ component of a constant external magnetic field, which is non-zero only in the region of the length  $ D_{0} $ between two TLL electrodes. Obviously, bosonic operators in equation~(\ref{4}) have the following symmetry: $ \hat{X}_{\text{L(R)}}^{+}=\hat{X}_{\text{R(L)}} $ as well as
\begin{equation} \label{5}
\left( \hat{X}_{\text{L(R)}}\Leftrightarrow \hat{X}_{\text{R(L)}}\right) = \left( \hat{y}\Leftrightarrow -\hat{y} \right).
\end{equation}
This symmetry, as it will be clear herein below, fixes definite values of relative Aharonov-Bohm phase of the tunneling electron in each vibronic channel. 
Furthermore, as it was mentioned in \cite{KG1}, since the ``charge''-density $\Phi_{+}(x)$-channel is decoupled at the Toulouse point, one can rewrite the current operator by means of  the ``current''-density channel $\Phi_{-}(x)$ only
\begin{equation} \label{6}
\hat{I}(\infty) =G_{0}\left[\hat{\Psi}_{-}^{+}\hat{\Psi}_{-}(-\infty)  -  \hat{\Psi}_{-}^{+}\hat{\Psi}_{-}(+\infty)\right],
\end{equation}
where $ G_{0}=e^{2}/h $ is the conductance quantum. Now, to solve the model of equations~(\ref{1})--(\ref{6}), a well-known quantum equation of motion (QEM) method can be used. The Heisenberg equations for fermionic operators take the form (at $ \Delta=0 $)
\begin{equation} \label{7}
\ri\hbar\partial_{t}\hat{d}=\gamma_{\text L}\hat{X}_{\text L}^{+} \hat{\Psi}_{-}(0)+\gamma_{\text R}\hat{X}_{\text R}^{+} \hat{\Psi}_{-}^{+}(0),
\end{equation}
\begin{equation} \label{8}
\ri\hbar\partial_{t}\hat{\Psi}_{-}(x)=-\ri\partial_{x}\hat{\Psi}_{-}(x)+\delta(x)\big(\gamma_{\text L}\hat{X}_{\text L}\hat{d}-\gamma_{\text R}\hat{X}_{\text R}^{+}\hat{d}^{+}\big),
\end{equation}
where $\delta(x)$ is the delta function. Integrating equation~(\ref{8}) in a small vicinity of the point $x=0$, one obtains
\begin{equation} \label{9}
\ri\left[\hat{\Psi}_{-}(0^{+})-\hat{\Psi}_{-}(0^{-})\right]=\gamma_{\text L}\hat{X}_{\text L}\hat{d}-\gamma_{\text R}\hat{X}_{\text R}^{+}\hat{d}^{+}.
\end{equation}
In the absence of magnetopolaronic coupling ($\hat{X}_{\text{L(R)}}^{+}=\hat{X}_{\text{L(R)}}=1$), equations~(\ref{7})--(\ref{9}) are reduced to equation~(\ref{6}) from reference~\cite{KG1}. Integrating formally equation~(\ref{7}) (similarly to reference~\cite{MY}) and substituting the obtained solution into equation~(\ref{9}), one can derive the following basic integral operator equation in the form of the operator-valued boundary condition at the physical point $ x=0 $: 
\begin{eqnarray}
&&\hbar\left\lbrace\hat{\Psi}_{-}(0^{+};t)-\hat{\Psi}_{-}(0^{-};t)\right\rbrace
\nonumber \\
&&=- \lim_{\alpha \rightarrow 0}\int_{0}^{t}\rd t'\Big\{\left[\gamma_{\text L}^{2}\hat{X}_{\text L}(t)\hat{X}_{\text L}^{+}(t')\hat{\Psi}_{-}(0;t')
+\gamma_{\text L}\gamma_{\text R}\hat{X}^{+}_{\text R}(t)\hat{X}^{+}_{\text R}(t')\hat{\Psi}_{-}^{+}(0;t')\right]
\re^{-\alpha(t-t')/\hbar}
\nonumber\\
&&+\left[\gamma_{\text R}^{2}\hat{X}_{\text R}^{+}(t)\hat{X}_{\text R}(t')\hat{\Psi}_{-}(0;t')
+\gamma_{\text L}\gamma_{\text R}\hat{X}_{\text L}^{+}(t)\hat{X}_{\text L}^{+}(t')\hat{\Psi}_{-}^{+}(0;t')\right] \re^{-\alpha(t-t')/\hbar}\Big\}.
\label{10}
\end{eqnarray}
Central operator equation (\ref{10}) should be complemented by equation for bosonic operator $\hat{y}$. The corresponding Heisenberg equation could be rewritten in the form of the Newton-like equation of motion for operator $\hat{y}$ with a quantum analog of Lorentz force in the right-hand side of the equation:
\begin{equation} \label{11}
\big(\partial_{t}^{2}+\omega_{0}^{2}\big)\hat{y}=-\phi\omega_{0}\hat{I}(0)=\phi\omega_{0}\left[\hat{\Psi}_{-}^{+}\hat{\Psi}_{-}(0^{+})  -  \hat{\Psi}_{-}^{+}\hat{\Psi}_{-}(0^{-})\right].
\end{equation}
Here, $ \hat{I}(0) $ denotes the ``MRL-current'' operator at the physical point $ x=0 $, i.e., at the boundary with a quantum dot. 

Now, if the magnetopolaronic coupling in the r.h.s. of equation~(\ref{11}) is small (assuming that $ \phi < 1 $ and, correspondingly, $ \phi^{2} \ll 1 $) then, to solve a problem of equations~(\ref{10}), (\ref{11}), one can neglect the r.h.s. of equation~(\ref{11}), and this corresponds to a more wide assumption regarding the fermion-boson factorization of all averages.\footnote{Actually, it is possible to show that fermion-boson factorization maintains in this magnetopolaronic Majorana-resonant-level model at arbitrary tunnel- and magnetopolaronic couplings, meaning that the latter magnetopolaronic MRL-model is \textit{exactly solvable}. However, a detailed analysis of this interesting feature of the model is insufficient for the magnetopolaronic blockade effect being predicted in this paper (because such an effect can be obtained already within the approach $ \phi^{2} \ll 1 $, as it is shown here). Hence, we postpone a detailed discussion on magnetopolaronic Majorana-resonant level model in a general case to our subsequent publication on the subject.} Under this assumption (which is similar to the one from reference~\cite{MY}), one can replace all the products of nonlinear bosonic operators $ \hat{X}_{\text{L(R)}}^{+}(t')\hat{X}_{\text{L(R)}}(t) $ and $\hat{X}_{\text{R(L)}}^{+}(t')\hat{X}_{\text{R(L)}}^{+}(t)$ in the basic operator equation (\ref{10}) by corresponding averages with quadratic Hamiltonian of a decoupled quantum harmonic oscillator (\ref{2}). As a result, within the approach $ \phi^{2} \ll 1 $, equation~(\ref{11}) is fulfilled with its ``free'' solution of the form: $\hat{y}(t)=(\hat{b}_{0}^{+}\re^{\ri\omega_{0}t}+\hat{b}_{0}\re^{-\ri\omega_{0}t}) $  (bosonic operators $\hat{b}_{0}^{+}$~($\hat{b}_{0}$) describe the creation (annihilation) of a free vibron and fulfil a standard bosonic commutation relation $[\hat{b}_{0},\hat{b}_{0}^{+}]=1$). Thus, all \textit{normal} $ \langle\hat{X}_{\text{L(R)}}^{+}(t')\hat{X}_{\text{L(R)}}(t) \rangle$ and \textit{anomalous} $\langle\hat{X}_{\text{R(L)}}^{+}(t')\hat{X}^{+}_{\text{R(L)}}(t)\rangle$ bosonic averages will be defined by two \textit{different} infinite sums over the index $ l $ (the number of vibrons being emitted or absorbed by the subsystem of interacting electrons). In these infinite sums, each term will be  proportional to a definite function of $ l $ and of inverse temperature $ \beta=1/T $ (see, e.g., references~\cite{OUR1,MY}). Especially, one will have
\begin{equation}\label{12}
\big\langle\hat{X}_{\text{L(R)}}^{+}(t')\hat{X}_{\text{L(R)}}(t) \big\rangle=
\re^{-\phi^{2}(1+2n_{\text b})}\sum^{+\infty}_{l=-\infty}F_{l}(\beta)\re^{\ri l\omega_{0}(t' - t)} 
\end{equation}
and
\begin{equation}\label{13}
\big\langle\hat{X}_{\text{R(L)}}^{+}(t')\hat{X}^{+}_{\text{R(L)}}(t)\big\rangle=
\re^{-\phi^{2}(1+2n_{\text b})}\sum^{+\infty}_{l=-\infty}(-1)^{l}F_{l}(\beta)\re^{\ri l\omega_{0}(t' - t)},
\end{equation}
where for the functions $ F_{l}(\beta) $ one has  $ F_{l}(\beta)=I_{l}[2\phi^{2}\sqrt{n_{\text b}(1+n_{\text b})}]\re^{-\beta \hbar\omega_{0} l/2}$, and $I_{l}(z)$ is the Bessel function of $l$-th order of the imaginary argument, $ n_{\text b}=(\re^{\,\beta \hbar\omega_{0}}-1)^{-1} $ is the Bose-Einstein distribution function (for details see e.g., references~\cite{OUR1,MY}). Take notice of an important $ (-1)^{l} $ factor in the ``anomalous'' sum (\ref{13}). This factor is a fingerprint of \textit{phase coherence conservation} during the resonant tunneling and is the one being responsible for the qualitatively novel physical effect of \textit{anomalous} magnetopolaronic blockade being revealed in what follows. Remarkably, corresponding \textit{anomalous} (or ``superconducting'') averages of the type $\langle \hat{X}_{\text{R(L)}}^{+}(t')\hat{X}_{\text{R(L)}}^{+}(t)\rangle$ were absent in the previous polaronic  modification of MRL-model from reference~\cite{MY}. Such anomalous averages describe the spatially non-local magnetopolaron-assisted Andreev-like tunneling (analogous to Andreev reflection of the corresponding spatially non-local composite fermion in the vicinity of a quantum dot from references~\cite{MY,KG1}). As for a fermionic part of the problem, following the method of references~\cite{KG1,MY}, at resonance by gate voltage (i.e., at $ \Delta=0 $), one can rewrite the decomposition for $\hat{\Psi}_{-}(x,t)$ fermionic operators from reference~\cite{KG1}: 
\begin{equation} \label{14}
\hat{\Psi}_{-}(x;t)=\int\frac{\rd k}{2\piup}\re^{\ri k(t-x)}\left\{
\begin{array}{ll}
\hat{a}_{k}\,, & x<0,  \\
\hat{b}_{k}\,, & x>0,
\end{array}
\right.
\end{equation}
where $\hat{a}_{k}^{+}$ ($\hat{a}_{k}$) are the standard fermionic creation (annihilation) operators. [Note that $ \hat{b}_{k}=t(k)\hat{a}_{k} $, where $ t(k) $ is the transmission amplitude.] Hence, similarly to references~\cite{KG1,MY}, using equations~(\ref{6}), (\ref{11})--(\ref{14}), one can write down the Landauer-type transport formula for the average current in the magnetopolaronic MRL-model: 
\begin{equation}\label{15}
\bar{I}(eV)=G_{0}\int \rd\varepsilon R_{\phi0}(\varepsilon)[n_{\text F}(\varepsilon-\mu_{\text L})-n_{\text F}(\varepsilon - \mu_{\text R})],
\end{equation}
where $ \mu_{\text L}-\mu_{\text R}=eV $ and $R_{\phi0}(\varepsilon)=1-|t(\varepsilon)|^{2}$ is the energy-dependent coefficient of ``Andreev-like'' reflection of $\hat{\Psi}_{-}$-fermions, which determines the transmission coefficient for physical electrons, and $ n_{\text F}(\varepsilon)=(\re^{\beta\varepsilon}+1)^{-1} $ is the Fermi-Dirac distribution function ($ \beta^{-1}=T $ is the temperature). As usual, at $ T \rightarrow 0 $, one has $n_{\text F}(\varepsilon - \mu_{\text{L(R)}})\rightarrow \theta(\varepsilon - \mu_{\text{L(R)}})  $ and for the differential conductance of the system $  G(eV)=\frac{\rd\bar{I}(eV)}{\rd V}$, one can write at $ T=0 $ from equation~(\ref{15}) $ G(eV)=G_{0}R_{\phi0}(eV) $. Starting from here, it is reasonable to consider the most distinct realization of the effect\footnote{For asymmetric case the effect will be qualitatively the same although it will be less distinct quantitatively.} for symmetric tunnel couplings $\gamma_{\text L}=\gamma_{\text R}=\sqrt{\Gamma_{0}}/2$.  
Since it turns out that zero-temperature ``off-set'' regime in the bias voltage: $ \mu_{\text L},\mu_{\text R},\Gamma_{0}\gg \hbar \omega_{0} \rightarrow 0 $ ($ T< \hbar\omega_{0} \neq 0 $) produces the most robust measurable effect, in what follows we restrict ourselves only to that limit of more general formulae. In the latter case, all virtual vibronic channels  contribute corresponding infinite sums in bosonic averages of equations~(\ref{12}), (\ref{13}), but due to  equations~(\ref{4}), (\ref{5}) and fermion-boson factorization approach, one can easily sum up these infinite sums over virtual vibronic channels. Indeed, as it follows from equations~(\ref{4}), (\ref{5}) (see also a similar derivation in reference~\cite{OUR1}), at $ t'=t $ or in the limit $ \omega_{0} \rightarrow 0$ (which concerns us here), the following (unitarity) condition: $ \sum^{+\infty}_{-\infty}F_{l}(\beta)=\re^{\phi^{2}} $ should fulfil for ``normal'' bosonic averages $ \langle\hat{X}_{\text{L(R)}}^{+}(t)\hat{X}_{\text{L(R)}}(t) \rangle$ [since in the low-temperature limit $ T< \hbar\omega_{0} $, one can put $ n_{\text b} \approx 0$ and hence we have $ \re^{-\phi^{2}(1+2n_{\text b})} \approx \re^{-\phi^{2}} $].  However, under the same approach ($ T< \hbar\omega_{0}\neq 0 $ and $ \hbar \omega_{0}\rightarrow 0 $) for the ``anomalous'' bosonic averages $\langle\hat{X}_{\text{R(L)}}^{+}(t')\hat{X}^{+}_{\text{R(L)}}(t)\rangle$ in equation~(\ref{10}) due to prefactors $ (-1)^{l} $ in the corresponding infinite sums, one has $\langle\hat{X}_{\text{R(L)}}^{+}(t')\hat{X}^{+}_{\text{R(L)}}(t)\rangle=\re^{-2\phi^{2}}$, which, obviously, ``breaks'' the unitarity of the corresponding Andreev-like resonant tunneling processes [mathematically, the latter occurs because in the zero-temperature limit for the corresponding anomalous sums, one has $ \sum^{+\infty}_{-\infty}(-1)^{l}F_{l}(\beta)\approx \sum^{+\infty}_{0}(-1)^{l}(\phi^{2})^{l}/l!=\re^{-\phi^{2}} $ and, hence, $ \re^{-\phi^{2}}\sum^{+\infty}_{-\infty}(-1)^{l}F_{l}(\beta)=\re^{-2\phi^{2}} $].

Thus, in the limits:  $ \phi^{2} \ll 1 $ and $ \hbar\omega_{0} \rightarrow 0 $, integrating the basic operator equation (\ref{10}), performing the mentioned bosonic summation  and solving the resulting algebraic equation together with its hermitian-conjugated equation with respect to $ t(\varepsilon) $, in the spirit of references~\cite{MY,KG1}, one can derive the following non-trivial formula for the \textit{effective} transmission coefficient of interacting electrons through the \textit{magnetopolaron coupled} Majorana-resonant level in the case of symmetric tunnel coupling
\begin{equation}\label{16}
R_{\phi0}(\varepsilon)=\frac{\big(4\Gamma^{2}_{0}\re^{-4\phi^{2}}\big)\varepsilon^{2}}{\varepsilon^{4}+2\Gamma^{2}_{0}\big(1+\re^{-4\phi^{2}}\big)\varepsilon^{2}+\Gamma^{4}_{0}\big(1-\re^{-4\phi^{2}}\big)^{2}}.
\end{equation}  

\begin{figure}[!t]
	\centering\includegraphics[width=0.59\textwidth]{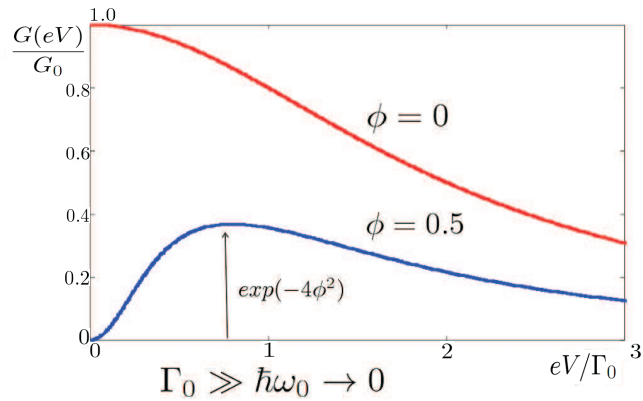}
	\vspace{-3mm}
	\caption{(Colour online) Zero-temperature differential conductance (in the units of conductance quantum $ G_{0} $) for the effective transmission coefficient from equation~(\ref{16}) in the limit: $ \varepsilon,\Gamma_{0}\gg \hbar\omega_{0} \rightarrow 0$ (while $ \hbar\omega_{0}\neq 0 $), as a function of bias voltage (in the units of $ \Gamma_{0} $) for two cases: i) $ \phi=0$ (red solid line); ii) $ \phi =0.5$ (blue solid line).}\label{fig1}
\vspace{-1mm}
\end{figure}

Then, substituting equation~(\ref{16}) into equation~(\ref{15}) in the limit $ T=0 $ and differentiating it with respect to $ V $, one arrives at the corresponding formula for zero-temperature differential conductance of the system (see figure~\ref{fig1}). First, at $ \phi=0$ and $\hbar\omega_{0}=0 $, formula (\ref{16}) results in: $ R_{0}= \Gamma^{2}_{0}/(\varepsilon^{2}+\Gamma^{2}_{0})$ a well-known Breit-Wigner transmission coefficient for resonant tunneling through $ g=1/2 $ TLLRL model with symmetric tunnel couplings \cite{KF,KG1}. Remarkably, in the lowest order in the small parameter $ \phi^{2}\ll 1 $, the above formula (\ref{16}) totally coincides (up to redefinition of $ \Gamma_{0} $ and $ \phi $) with formula (\ref{16}) from reference~\cite{MK} for the effective transmission coefficient $ D^{\text{eff}}_{0}(\omega) $, which was calculated for the polaronic MRL-model in the limits of perturbation theory in small constant of electromechanical coupling. 

In figure~\ref{fig1}, the zero-temperature differential conductance with transmission coefficient of equation~(\ref{16}) is plotted for two limiting cases: i) when $ \phi=0$ (red solid line) and ii) when $ \phi\neq 0$, ($ \phi=0.5$, blue solid line). From figure~\ref{fig1}, one can see a sharp difference between these two cases. Indeed, in the case $ \phi=0$, we have a usual Lorenzian-shape curve for transmission, while in the case $ \phi\neq 0$ (even at $ \phi^{2}\ll 1 $), the \textit{destructive interference} between different virtual vibronic channels of electron tunneling results in a strongly nonmonotonous behaviour of the effective transmission (\ref{16}) and respective zero-temperature differential conductance (blue curve in figure~\ref{fig1}). As a result, the zero-temperature differential conductance in the case of magnetopolaronic MRL-model at $ eV \neq 0$ reaches its maximal value [being equal to $ G_{0}\exp(-4\phi^{2}) $] at non-zero bias voltage value (which is approximately equal to $ \Gamma_{0} $ in such a case, where $ \phi^{2} \ll 1$). Whereas at $ eV=0 $ (i.e., at $ \mu_{\text L}=\mu_{\text R} \gg 0 $), the transmission coefficient $ R_{\phi0}(0) $ tends to zero, contrary to the situation without any quantum vibrations ($ \phi=0$), while at any $ \varepsilon, eV \gg \Gamma_{0} $ it decreases smoothly, similarly to the case $ \phi=0$. This fact means that in the magnetopolaronic MRL-system at zero temperature, in the limit: $ \Gamma_{0}\gg \hbar\omega_{0} \rightarrow 0 $, it is impossible to ``compensate'' the magnetopolaronic Franck-Condon blockade (see \cite{KO}) of a resonant tunneling \cite{OUR1} even by means of a very high bias voltage at $ eV \gg \hbar\omega_{0} $. 

Qualitatively, a fermion-boson decoupling in the equations~(\ref{10})--(\ref{13}) means that in the magnetopolaronic MRLM case, a vibronic subsystem has a complete set of eigenstates, which are just eigenstates of a free quantum harmonic oscillator. Thus, each vibronic eigenfunction (i.e., each eigenfunction of a magnetopolaron) should have a certain fixed parity which should \textit{conserve} in the resonant tunneling process with respect to symmetry transformations of equation~(\ref{5}). Such a parity is equal to $ (-1)^{l} $, where $ \vert l \vert $ is the number of nodes of the $l$-th eigenfunction of a quantum oscillator (or equally, the number of vibrons in the corresponding vibronic eigenstate). 

These fixed parities' factors could be explained as the ones being ``kept'' by definite fixed values of the Aharonov-Bohm (A-B) phases $ \Delta\varphi^{\text{odd/even}}_{l}= \piup\vert l \vert $ in each $ l $-th vibronic channel of tunneling (where $ l $ can be either odd or even number). Thus, resonant ``Andreev-like'' magnetopolaronic tunneling in the MRL-model turns out to be driven by the values $ \pm \piup $ of Aharonov-Bohm phase differences between any $ l $-th and $ l' $-th vibronic channels  of magnetopolaron tunneling. One could suppose that in the magnetopolaronic MRL-model, such fixed values of magnetic A-B phase in each $ l $-th channel of tunneling play the role of a Berry phase \cite{DX1,DX2}, which is acquired by a real electron in the respective virtual vibronic channel of tunneling. This phase-coherent effect is the one which ``drives'' the magnetopolaronic MRL-system to the above described situation of anomalous magnetopolaronic blockade of differential conductance (plotted in figure~\ref{fig1}) in the zero-temperature limit, even at small values of a magnetopolaronic coupling constant and even in the case of symmetric tunnel coupling.

In conclusion, a novel formula was obtained for a transmission coefficient and respective average current of strongly interacting electrons in the Majorana-resonant level magnetopolaronic model. As a result, it was found that in the zero-temperature limit, even in the case of weak  magneto-mechanical coupling, it is impossible to compensate the effect of strong \textit{anomalous} magnetopolaronic blockade by means of a bias voltage in the case where the vibron energy is the smallest (nonzero) energy scale in the system (except the temperature). In principle, this qualitatively new effect, being predicted theoretically in the above, could be measured experimentally in single-electron transistors with one ``short'' suspended and quantum-vibrating in the transverse magnetic field carbon nanotube in the role of a quantum dot with its one resonant level coupled by means of two tunnel contacts to two proper (and long enough) carbon nanotubes in the role of the corresponding one-dimensional $ g=1/2 $ Luttinger liquid leads. One could use the described effect of anomalous magnetopolaronic blockade in MRL-model in order to detect a possible MRL-type of tunnel coupling in SETs as well as for the estimations of ultra-small eigenfrequencies and the corresponding zero-point-oscillation amplitudes in a variety of carbon nanotubes-based MRL magnetopolaronic quantum mesoscopic systems.   

\section*{Acknowledgements}

The author thanks A.~Komnik, I.V.~Krive, S.I.~Kulinich, R.I.~Shekhter, F.~Pistolesi, Y.~Gefen and K.~Shtengel  for valuable discussions.

\newpage
\ukrainianpart

\title{Аномальна магнітополяронна блокада резонансного електронного 
 тунелювання в одноелектронних транзисторах з Майоранівським резонансним рівнем 
 при нульовій температурі}
\author{Г.О. Скоробагатько}
\address{
Інститут фізики конденсованих систем НАН України, вул. Свєнціцького,  1, 79011  Львів,  Україна
}

\makeukrtitle

\begin{abstract}
\tolerance=3000%
Розглянуто магнітополяронне узагальнення моделі Майоранівського резонансного 
 рівня для однорівневого вібруючого (у зовнішньому перпендикулярному магнітному 
 полі) квантового доту, який під'єднано до двох довгих (напів-безкінечних) 
 Латинжерівських електродів з латинжерівським кореляційним параметром, що 
 дорівнює 1/2. У наближенні ферміон-бозонної факторизації відповідних середніх 
 отримано якісно нову, нетривіальну формулу для ефективного коефіцієнту 
 проходження електронів у випадку їх резонансного магнітополяронного 
 тунелювання через резонансний Майоранівський рівень доту з квантовими 
 вібраціями, а також відповідний вираз для диференційного кондактансу 
 розглянутої системи. Використане в роботі наближення ферміон-бозонної 
 факторизації середніх має сенс у випадку слабкої магнітополяронної (або 
 інакше, магніто-механічної) взаємодії у системі. Натомість, в цій роботі було 
 виявлено, що, незважаючи на використану умову слабкості взаємодії між 
 ферміонною та бозонною підсистемами, сильнокорельований електронний транспорт 
 у розглянутій специфічній моделі демонструє риси сильної (тож, відповідно, 
 ``аномальної'') магнітополяронної блокади при нульовій температурі, якщо енергія 
 кванту механічної вібрації (або віброну) є найменшим (але ненульовим) 
 енергетичним параметром системи. Такий ефект слід інтерпретувати як фазовокогерентне магнітополяронне резонансне тунелювання Андріївського типу, яке 
 витікає із специфічної, Майоранівської симетрії розглянутого тунельного 
 гамільтоніану з магнітополяронним зв'язком. Ефект, передбачений в цій роботі, 
 може бути використаний як маркер експериментальної реалізації Майоранівського 
 резонансного рівня в одноелектронних транзисторах, а також як спосіб 
 експериментального вимірювання ультра-слабких нульових коливань підвішених 
 вуглецевих нанотрубок при тунелюванні крізь них електронів у режимі 
 Майоранівського резонансного рівня відповідного одноелектронного транзистора.
\keywords електронне тунелювання, модель  Майоранівського  резонансного рівня, рідина  Латинжера, реферміонізація, магніто-механічний  зв'язок, магнітополярон

\end{abstract}


\begin{thebibliography}{99}
\bibitem{PP} Park H., Park J., Lim A.K.L., Anderson E.H., Alivisatos A.P., McEuen P.L.,   Nature, 2000, {\bf 407}, 57, \doi{10.1038/35024031}.
\bibitem{PT} Postma H.W.Ch., Teepen T., Yao Z., Grifoni M., Dekker C., Science, 2001, {\bf 293}, 76,\\ \doi{10.1126/science.1061797}.
\bibitem{NR} Nitzan A.,  Ratner M.A.,  Science, 2003, {\bf 300}, 1384, \doi{10.1126/science.1081572}.
\bibitem{GR} Galperin M., Ratner M.A.,  Nitzan A.,  J. Phys.: Condens. Matter, 2007, {\bf 19}, 103201,\\ \doi{10.1088/0953-8984/19/10/103201}.
\bibitem{GS}  Glazman L.I.,  Shekhter R.I.,  Zh. Eksp. Teor. Fiz., 1988, {\bf 94}, 292 (in Russian), [Sov. Phys. JETP, 1988, {\bf 67}, 163]. 
\bibitem{FL} Braig S.,  Flensberg K., Phys. Rev. B, 2003, {\bf 68}, 205324, \doi{10.1103/PhysRevB.68.205324}.
\bibitem{MCK} Lundin U.,  McKenzie R.H.,  Phys. Rev. B, 2002, {\bf 66}, 075303, \doi{10.1103/PhysRevB.66.075303}.
\bibitem{NEW} Maier S., Schmidt T.L.,  Komnik A.,  Phys. Rev. B, 2011, {\bf 83}, 085401, \doi{10.1103/PhysRevB.83.085401}.
\bibitem{KF} Kane C.L.,  Fisher M.P.A.,  Phys. Rev. B, 1992, {\bf 46}, 15233, \doi{10.1103/PhysRevB.46.15233}.
\bibitem{MY} Skorobagatko G.A., Phys. Rev. B, 2012, {\bf 85}, 075310, \doi{10.1103/PhysRevB.85.075310}.
\bibitem{KG1} Komnik  A., Gogolin A.O.,  Phys. Rev. Lett., 2003, {\bf 90}, 246403, \doi{10.1103/PhysRevLett.90.246403}.
\bibitem{KG2}  Komnik A., Gogolin A.O., Phys. Rev. B, 2003, {\bf 68}, 235323, \doi{10.1103/PhysRevB.68.235323}.
\bibitem{OUR1} Skorobagatko G.A., Kulinich S.I., Krive I.V., Shekhter R.I.,  Jonson M., Low Temp. Phys., 2011, {\bf 37}, 1032, \doi{10.1063/1.3674185}.
\bibitem{Pist} Rastelli G., Houzet M., Glazman L., Pistolesi F.,  C.R. Phys., 2012, {\bf 13}, 410, \doi{10.1016/j.crhy.2012.03.001}.
\bibitem{KR} Krive I.V., Palevski A., Shekhter R.I., Jonson M., Low Temp. Phys., 2010, {\bf 36}, 119, \doi{10.1063/1.3319350}.
\bibitem{AK} Komnik A.,   Phys. Rev. B, 2009, {\bf 79}, 245102, \doi{10.1103/PhysRevB.79.245102}.
\bibitem{NG} Nazarov Yu.V.,  Glazman L.I., Phys. Rev. Lett., 2003, {\bf 91}, 126804, \doi{10.1103/PhysRevLett.91.126804}.
\bibitem{MK} Maier S., Komnik A., Phys. Rev. B, 2010, {\bf 82}, 165116, \doi{10.1103/PhysRevB.82.165116}.
\bibitem{KO} Koch J., von Oppen F.,  Andreev A.V.,  Phys. Rev. B, 2006, {\bf 74}, 205438, \doi{10.1103/PhysRevB.74.205438}.
\bibitem{DX1} Xiao D., Chang M.-C., Niu Q.,   Rev. Mod. Phys., 2010, {\bf 82}, 1959, \doi{10.1103/RevModPhys.82.1959}. 
\bibitem{DX2} Whitney R.S., Gefen Y., In: Electronic Correlations: from Meso- to Nanophysics, Martin T., Montambaux G., Tr\^{a}n Thanh~V\^{a}n~J.~(Eds.), EDP Sciences, Paris, 2001, 291--298. 




\end{thebibliography}
\end{document}